\documentstyle[sprocl]{article}

\input{epsf}
\def\be{\begin{equation}}
\def\ee{\end{equation}}
\def\bea{\begin{eqnarray}}
\def\eea{\end{eqnarray}}
\begin{document}

\title{THE SECOND SUPERSTRING REVOLUTION}

\author{JOHN H. SCHWARZ}

\address{California Institute of Technology, Pasadena, Calif.  91125, USA}

\maketitle\abstracts{
Recent developments in superstring theory have led to major advances in
understanding.  After reviewing where things stood earlier, we sketch the
impact of recently identified dualities and D-branes, as well as a hidden
eleventh dimension.}

It is my honor and pleasure to speak at this conference in memory of Andrei
Sakharov on the occasion of his 75th birthday.  Even though I never met with
Sakharov personally, I understand from my good friend Professor Fradkin that
Sakharov expressed considerable interest in superstring theory during his exile
in Gorky.  I would have enjoyed discussing the subject with him, but after his
return to Moscow he had other more urgent matters to deal with.  I admired him
for his courage and persistent efforts on behalf of human rights and
disarmament.

Major advances in understanding of the physical world have been achieved during
the past century by focusing on apparent contradictions between
well-established theoretical structures.  In each case the reconciliation
required a better theory, often involving radical new concepts and striking
experimental predictions.  Four major advances of this type are indicated in
Figure 1.\cite{strominger1}  These advances were the discoveries of special
relativity, quantum mechanics, general relativity, and quantum field theory.
This was quite an achievement for one century, but it leaves us with one
fundamental contradiction that still needs to be resolved, namely the clash
between general relativity and quantum field theory.  Many theorists are
convinced that superstring theory will provide the answer.  There have been
major advances in our understanding of this subject, which I consider to
constitute the ``second superstring revolution,'' during the past two years.
The plan for this brief report is to sketch where things stood after the first
superstring revolution (1984-85) and then to describe the recent developments
and their implications.

\begin{figure}[t]
\centerline{\epsfxsize=4truein \epsfbox{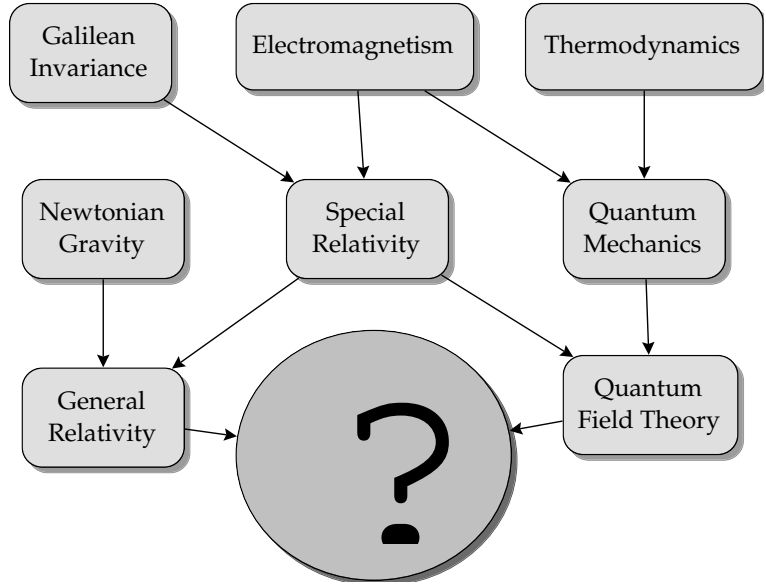} }
\caption{Contradictions lead to better theories.}
\end{figure}

There are various problems that arise when one attempts to combine general
relativity and quantum field theory.  The field theorist would point to the
breakdown of renormalizability -- the fact that short-distance singularities
become so severe that the usual methods for dealing with them no longer work.
By replacing point-like particles with one-dimensional extended strings, as the
fundamental objects, superstring theory certainly overcomes the problem of
perturbative non-renormalizability.  A relativist might point to a different
set of problems including the issue of how to understand the causal structure
of space-time when the metric has quantum-mechanical excitations.  There are
also a host of problems associated to black holes such as the fundamental
origin of their thermodynamic properties and an apparent loss of quantum
coherence.  The latter, if true, would require a modification in the basic
structure of quantum mechanics.  The relativist's set of issues cannot be
addressed
properly in a perturbative setup, but the recent discoveries are leading to
non-perturbative understandings that should help in addressing them.
Most string theorists expect that the
theory will provide satisfying resolutions of these problems without any
revision in the basic structure of quantum mechanics.  Indeed, there are
indications that someday quantum mechanics will be viewed as an implication
(or at least a necessary ingredient) of superstring theory.

When a new theoretical edifice is proposed, it is very desirable to identify
distinctive testable experimental predictions.  In the case of superstring
theory there have been no detailed computations of the properties of elementary
particles or the structure of the universe that are convincing, though many
valiant attempts have been made.
In my opinion, success in such enterprises requires a
better understanding of the theory than has been achieved as yet.  It is very
difficult to assess whether this level of understanding is just around the
corner or whether it will take many decades and several more revolutions.  In
the absence of this kind of confirmation, we can point to three general
``predictions'' of superstring theory that are very encouraging.  The first is
the existence of gravitation, approximated at low energies by general
relativity.  No other quantum theory can claim to have done this (and I suspect
that no other ever will).  The second is the fact that superstring vacua
generally
include Yang--Mills gauge theories like those that make up the ``standard
model'' of elementary particles.  The third general prediction, not yet
confirmed experimentally, is the existence of supersymmetry at low energies
(the electroweak scale).  There are tantalizing hints that it may show up this
century at CERN or Fermilab.

The history of string theory is very fascinating, with many bizarre twists and
turns.  It has not yet received the attention it deserves from historians of
science.  Here we will settle for a very quick sketch.  The subject arose in
the late 1960's in an attempt to describe strong nuclear forces.  It was a
quite active subject for about five years, but it ran into theoretical
difficulties, and QCD came along as a convincing theory of the strong
interaction.  During this period (in 1971) it was discovered
that the inclusion of fermions requires
world-sheet supersymmetry.\cite{ramond}  This led to the development of
space-time supersymmetry, which was eventually recognized to be a generic
feature of consistent string theories (hence the name ``superstrings'').  In
1974 Jo\"el Scherk and I proposed that the problems of string theory could be
turned into virtues  if it were used as a framework for realizing Einstein's
old dream of ``unification,'' rather than as a theory of hadrons.\cite{scherk}
Specifically, we pointed out that it would provide a perturbatively finite
theory that incorporates general relativity.  One implication of this change in
viewpoint was that the characteristic size of a string became the Planck length
$L_{PL} = (\hbar G/c^3)^{1/2} \sim 10^{-33} cm$, some 20 orders of magnitude
smaller than previously envisaged.  (More refined analyses lead to a string
scale $L_{ST}$ that is about two orders of magnitude larger than the Planck
length.)  In any case, experiments at existing accelerators cannot resolve
distances shorter than about $10^{-16} cm$, which explains why the
point-particle approximation of ordinary quantum field theories is so
successful.

In 1984-85 there was a series of discoveries~\cite{green} that convinced many
theorists that superstring theory is a very promising approach to unification.
 Almost overnight, the subject was transformed from an intellectual backwater
to one of the most active areas of theoretical physics, which it has remained
ever since.  By the time the dust settled, it was clear that there are five
different superstring theories, each requiring ten dimensions (nine space and
one time), and that each has a consistent perturbation expansion.
The five theories, about which I'll say more later, are denoted type I, type
IIA, type IIB, $E_8 \times E_8$ heterotic (HE, for short), and SO(32) heterotic
(HO, for short).  The type II theories have two supersymmetries in the
ten-dimensional sense, while the other three have just one.  The type I theory
is special in that it is based on unoriented open and closed strings, whereas
the other four are based on oriented closed strings.
%The IIA theory is special in that it is non-chiral (i.e., it is parity
% conserving),
%whereas the other four are chiral (parity violating).

A string's space-time history is described by functions $x^\mu (\sigma,\tau)$,
which map the string's two-dimensional ``world sheet'' $(\sigma,\tau)$ into
space-time $x^\mu$.  There are also other world-sheet fields that describe
other degrees of freedom, such as those associated with supersymmetry and gauge
symmetries.  Surprisingly, {\em classical} string theory dynamics is described
by
a conformally invariant 2D {\em quantum} field theory
\begin{equation}
S = (1/L_{ST})^2 \int d \sigma d \tau {\cal L} (x^\mu (\sigma, \tau), \ldots) .
\end{equation}
What distinguishes one-dimensional strings from higher
dimensional analogs is the fact that this 2D theory is renormalizable.
(Objects with $p$ dimensions, called ``$p$-branes," have a $(p+1)$-dimensional
world volume.)  Perturbative quantum string theory can be formulated
by the Feynman sum-over-histories method.  This
amounts to associating a genus $h$ Riemann surface to an $h$-loop string theory
Feynman diagram.  The attractive features of this approach are that there is
just one diagram at each order of the perturbation expansion and that each
diagram represents an elegant (though complicated) mathematical expression that
is ultraviolet finite.  The main drawback of this approach is that it gives no
insight into how to go beyond perturbation theory.

To have a chance of being realistic,
the six extra space dimensions must curl up into a tiny geometrical space,
whose
size should be comparable to $L_{ST}$.  Since space-time geometry is
determined dynamically (as in general relativity) only geometries that satisfy
the dynamical equations are allowed.
The HE string theory, compactified on a particular kind
of six-dimensional space called a Calabi--Yau manifold has many qualitative
features at low energies that resemble the standard model.  In particular, the
low mass fermions occur in families, whose number is controlled by the
topology of the CY manifold.  These successes have been achieved in a
perturbative framework, and are necessarily qualitative at best, since
non-perturbative phenomena are essential to an understanding of supersymmetry
breaking and other important matters of detail.

The second superstring revolution (1994-??) has brought non-perturbative string
physics within reach.  The key discoveries were the recognition of amazing and
surprising ``dualities.''  They have taught us that what we viewed previously
as five
theories is in fact five different perturbative expansions of a single
underlying theory about five different points!  It is now clear that there is a
unique theory, though it may allow many different vacua.  For example, a sixth
special vacuum involves an 11-dimensional Minkowski space-time. Another lesson
we have learned is that, non-perturbatively, objects of more than one dimension
(membranes and higher ``$p$-branes'') play a central role.  In most respects
they
appear just as fundamental as strings (which can now be called one-branes),
except that a perturbation expansion cannot be based on $p$-branes with $p >
1$.

Three kinds of dualities, called $S,T$, and $U$, have been identified.  It can
sometimes happen that theory $A$ at strong coupling is equivalent to theory $B$
at weak coupling, in which case they are said to be $S$ dual.  Similarly, if
theory $A$ compactified on a space of large volume is equivalent to theory $B$
compactified on a space of small volume, then they are called $T$ dual.
Combining these
ideas, if theory $A$ compactified on a space of large (or small) volume is
equivalent to
theory $B$ at strong (or weak) coupling, they are called $U$ dual.  If theories
$A$ and $B$ are the same, then the duality becomes a self-duality, and it can
be
viewed as a (gauge) symmetry. $T$ duality, unlike $S$ or $U$ duality, can be
understood perturbatively, and therefore it was discovered between the string
revolutions.

The basic idea of $T$ duality
(for a recent discussion see [5])  can be illustrated by
considering a compact dimension consisting of a circle of radius $R$.
In this case there are two kinds of excitations to consider.  The first, which
is not special to string theory, are Kaluza--Klein momentum excitations on the
circle, which contribute $(n/R)^2$ to the energy squared, where $n$ is an
integer.   Winding-mode excitations, due to
a closed string winding $m$ times around the circular dimension,
are special to string theory.  If $T = (2\pi
L_{ST}^2)^{-1}$ denotes the string tension (energy per unit length), the
contribution to the energy squared is $(2\pi R m T)^2$.  $T$ duality exchanges
these two kinds of excitations by mapping $m \leftrightarrow n$ and $R
\leftrightarrow L_{ST}^2 /R$.  This is part of an exact map between a $T$-dual
pair
$A$ and $B$.  One implication is that usual geometric concepts break down at
short distances, and classical geometry is replaced by ``quantum geometry,''
which is described mathematically by 2D conformal field theory. It also
suggests a generalization of the Heisenberg uncertainty  principle according
to which the best possible spatial resolution $\Delta x$ is bounded below not
only by the reciprocal of the momentum spread, $\Delta p$, but also by the
string
scale $L_{ST}$.  (Including non-perturbative effects, it may be possible to
do a little better and reach the Planck scale.)
Two important examples of  superstring theories that are $T$-dual when
compactified on a circle are the IIA and IIB theories and the HE and HO
theories.  These two dualities reduce the number of distinct theories from five
to
three.

Suppose now that a pair of theories ($A$ and $B$) are
$S$ dual. This means that if $f$ denotes any physical
observable and $\lambda$ denotes the coupling constant, then
$f_A (\lambda) = f_B (1/\lambda)$.  This duality, whose recognition
was the first step in the current revolution,\cite{font} generalizes the
electric-magnetic symmetry of Maxwell theory.  The point is that since the
Dirac
quantization condition implies that the basic unit of magnetic charge is
inversely
proportional to the unit of electric charge, their interchange
amounts to an inversion of the charge (which is the coupling constant).
$S$ duality relates the type I theory to the HO theory and the IIB theory to
itself.  This explains the strong coupling behavior of those three
theories. The understanding of how the IIA and HE theories behave at
strong coupling, which is by now well-established, came as quite a
surprise.  In each of these cases there is an 11th dimension that becomes large
at strong
coupling, the scaling law being $L_{11} \sim \lambda^{2/3}$.  In the IIA case
the 11th dimension is a circle, whereas in the HE case it is a line interval
(so that the eleven-dimensional space-time has two ten-dimensional boundaries).
 The strong coupling limit of either of these theories gives an 11-dimensional
Minkowski space-time.  The eleven-dimensional description of the underlying
theory is called ``M theory.''  As yet, it is less well understood than the
five
10-dimensional string theories.  The various connections among theories
that we've mentioned are sketched in Figure 2.  ($S^1$ denotes a circle and $I$
denotes a line interval.) There are many additional dualities that arise when
more dimensions are compactified, which will not be described here.

\begin{figure}[t]
\centerline{\epsfxsize=4truein \epsfbox{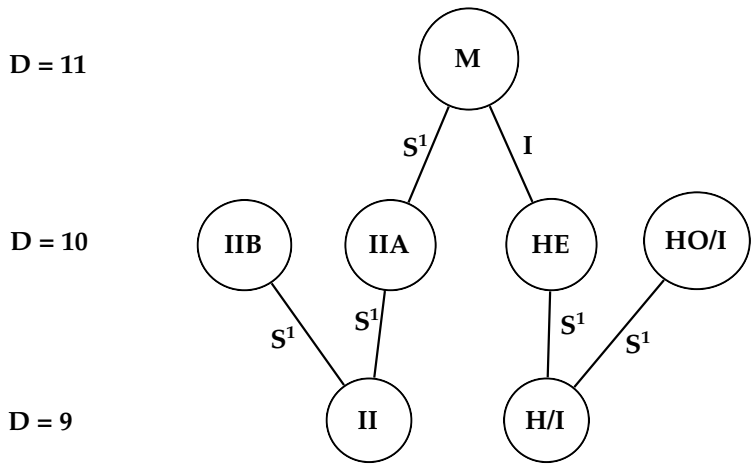} }
\caption{All superstring theories are connected.}
\end{figure}

Another source of insight into
non-perturbative properties of superstring theory has arisen from the study of
a
special class of $p$-branes called Dirichlet $p$-branes (or D-branes for
short).
The name derives from the boundary conditions assigned to the ends
of open strings.  The usual open strings of the type I theory have Neumann
boundary
conditions at their ends, but $T$ duality implies the existence of dual open
strings with Dirichlet boundary conditions in the dimensions that are
$T$-transformed.  More generally, in type II theories, one can consider open
strings with
\begin{equation}
\begin{array}{cll}
{\partial x^\mu\over\partial\sigma}|_{\sigma = 0} &= 0 & \mu = 0,1,\ldots,
p\\
&&\\
x^\mu|_{\sigma = 0} &= x_0^\mu & \mu = p + 1, \ldots, 9.
\end{array}
\end{equation}
At first sight this appears to break the Lorentz invariance of the theory,
which is paradoxical.  The resolution of the paradox is that strings end on
a $p$-dimensional dynamical object -- a D-brane.
D-branes have been studied for a number of years, but their significance was
explained by Polchinski only recently.~\cite{polchinski} They
are important because they make it possible to study the
excitations of the brane using the renormalizable 2D quantum field theory of
the open string instead of the non-renormalizable world-volume theory of the
D-brane itself.  In this way it becomes possible to compute non-perturbative
phenomena using perturbative methods!  Many (but not all) of the previously
identified $p$-branes are D-branes.  Others are related to D-branes by duality
symmetries so that they can also be brought under mathematical control.

D-branes have found many interesting applications, but the most remarkable of
these concerns the study of black holes.  Strominger and
Vafa~\cite{strominger2}
(and subsequently many others) have shown that D-brane techniques can be
used to  count the quantum microstates associated to classical black
hole configurations.  The simplest case, which was studied first, is static
extremal charged black holes in five dimensions.   Strominger and Vafa
showed that for large values of the charges
the entropy (defined by $S = \log N$, where $N$ is the number of quantum states
that
system can be in) agrees with the Bekenstein--Hawking prediction (1/4 the area
of the event horizon).  This result has been generalized to black holes in 4D
as
well as to ones that are near extremal (and radiate correctly) or rotating.  In
my
opinion, this is a truly dramatic advance.  It has not yet been proved
that there is no loss of quantum coherence, but I expect that result to
follow in due course.

I have touched on some of the highlights of the current revolution, but there
is much more that does not fit here.  For example, I have not discussed the
dramatic discoveries of Seiberg and Witten~\cite{seiberg} for supersymmetric
gauge theories and their extensions to string theory.  Another important
development due to Vafa~\cite{vafa} (called F theory) has made it possible to
construct large new classes of non-perturbative vacua of Type IIB superstrings.

Despite all the progress that has taken place in our understanding of
superstring theory, there are many important questions whose answers
are still unknown.  It is not clear how many
important discoveries still remain to be made before it will be possible to
answer the
ultimate question that we are striving towards -- why does the universe
behave the way it does?  Short of that, we have some other pretty big
questions.  What is the best way to formulate the theory?  How and why is
supersymmetry broken?  Why is the cosmological constant so small (or zero)?
How is a realistic vacuum chosen?  What are the cosmological implications of
the theory?  What testable predictions can we make? Stay tuned.

I am grateful to Patricia Schwarz for the figures.


\section*{References}
\begin{thebibliography}{99}

\bibitem{strominger1} A. Strominger, {\em Proceedings of the Sixth Marcel
Grossman Conference on General Relativity}, (June 1991) hep-th/9110011.

\bibitem{ramond} P. Ramond, {\em Phys. Rev.} {\bf D3} 2415 (1971); A. Neveu and
J.H. Schwarz, {\em Nucl. Phys.} {\bf B31} 86 (1971).

\bibitem{scherk} J. Scherk and J.H. Schwarz, {\em Nucl. Phys.} {\bf B81} 118
(1974).

\bibitem{green} M.B. Green and J.H. Schwarz, {\em Phys. Lett.} {\bf 149B} 117
(1984); D.J. Gross, J.A. Harvey, E. Martinec, and R. Rohm, {\em Phys. Rev.
Lett.}
{\bf 54} 502 (1985); P. Candelas, G.T. Horowitz, A. Strominger, and E. Witten,
{\em Nucl. Phys.} {\bf B258} 46 (1985).

\bibitem{witten} E. Witten, {\em ``Reflections on the Fate of Spacetime,''}
p. 24 in April 1996 Physics Today.

\bibitem{font} A. Font, L. Iba\~nez, D. L\"ust, and F. Quevedo, {\em Phys.
Lett.}
{\bf B249} 35 (1990); A. Sen, {\em Int. J. Mod. Phys.} {\bf A9} 3707 (1994);
{\em Phys. Lett.} {\bf B329} 217 (1994); C. Hull and P. Townsend, {\em Nucl.
Phys.} {\bf B438} 109 (1995); E. Witten, {\em Nucl. Phys.} {\bf B443} 85
(1995).

\bibitem{polchinski} J. Polchinski, {\em Phys. Rev. Lett.} {\bf 75} 4724
(1995).

\bibitem{strominger2} A. Strominger and C. Vafa, {\em ``Microscopic Origin of
Bekenstein--Hawking Entropy,''} hep-th/9601029.

\bibitem{seiberg} N. Seiberg and E. Witten, {\em Nucl. Phys.} {\bf B426} 19
(1994) and {\em Nucl. Phys.} {\bf B431} 484 (1994).

\bibitem{vafa} C. Vafa, {\em ``Evidence for F Theory,''} hep-th/9602022.

\end{thebibliography}
\end{document}